\documentclass[a4paper, aps, pra, reprint]{revtex4-1}
\usepackage[cm]{fullpage}
\usepackage[T1]{fontenc}
\usepackage{amssymb, amsmath, amsthm}
\makeatletter
\def\amsbb{\use@mathgroup \M@U \symAMSb}
\makeatother
\usepackage[mathscr]{euscript}
\usepackage{mathtools}
\usepackage{verbatim}
\usepackage{bbold}
\usepackage{xcolor}
\definecolor{darkred}{RGB}{200, 0, 0}
\definecolor{darkgreen}{RGB}{0, 100, 0}
\definecolor{darkblue}{RGB}{0, 0, 200}
\usepackage{hyperref}
\hypersetup{colorlinks, breaklinks, linkcolor=darkred, urlcolor=darkblue, citecolor=darkgreen}

\newcommand{\prlparagraph}[1]{\textsl{#1.}---}

\newcommand{\nbox}[2][9]{\hspace{#1pt} \mbox{#2} \hspace{#1pt}}

\DeclareMathOperator{\tr}{tr}

\def \diracspacing {0.7pt}
\newcommand{\ave}[1]{\langle #1 \rangle}
 
\newcommand{\ket}[1]{| \hspace{\diracspacing} #1 \rangle}

\newcommand{\ketbra}[2]{| \hspace{\diracspacing} #1 \rangle \langle #2 \hspace{\diracspacing} |} 
 
\newcommand{\ketbraq}[1]{\ketbra{#1}{#1}}

\newcommand{\norm}[2][]{#1| \! #1| #2 #1| \! #1|}

\newcommand{\cB}{\mathcal{B}}

\newcommand{\cQ}{\mathcal{Q}}

\newcommand{\cS}{\mathcal{S}}

\newtheorem{prop-glob}{Proposition}
\newcommand{\lmax}{\lambda_{\textnormal{max}}}
\newcommand{\trsym}{\cQ}
\newcommand{\q}{\sqrt{2}}
\newcommand{\alice}{Alessandro}
\newcommand{\bob}{Bruno}
\newcommand{\charlie}{Carol}
\begin{document}
\title{Analytic and nearly optimal self-testing bounds for the\\Clauser-Horne-Shimony-Holt and Mermin inequalities}
\author{J\k{e}drzej Kaniewski}
\email{jkaniewski@math.ku.dk}
\affiliation{Department of Mathematical Sciences, University of Copenhagen, Universitetsparken 5, 2100 Copenhagen, Denmark}
\date{\today}
\begin{abstract}
Self-testing refers to the phenomenon that certain extremal quantum correlations (almost) uniquely identify the quantum system under consideration. For instance observing the maximal violation of the CHSH inequality certifies that the two parties share a singlet. While self-testing results are known for several classes of states, in many cases they are only applicable if the observed statistics are almost perfect, which makes them unsuitable for practical applications. Practically relevant self-testing bounds are much less common and moreover they all result from a single numerical method (with one exception which we discuss in detail). In this work we present a new technique for proving analytic self-testing bounds of practically relevant robustness. We obtain improved bounds for the case of self-testing the singlet using the CHSH inequality (in particular we show that non-trivial fidelity with the singlet can be achieved as long as the violation exceeds $\beta^{*} = (16 + 14 \q)/17 \approx 2.11$). In case of self-testing the tripartite GHZ state using the Mermin inequality we derive a bound which not only improves on previously known results but turns out to be tight. We discuss other scenarios to which our technique can be immediately applied.
\end{abstract}
\pacs{03.67.-a}
\maketitle
\prlparagraph{Introduction}In 1964 John Bell showed that correlations resulting from any classical theory are restricted by certain constraints (now known as \emph{Bell inequalities})~\cite{bell64} and moreover that these might be violated by quantum systems. Nowadays Bell nonlocality is an active field with numerous applications~\cite{brunner14}. One of the most striking consequences of Bell's theorem is the fact that the non-classical nature of two (or more) devices can be verified by a classical user.
If we moreover assume quantum mechanics to be the underlying theory, we find that certain extremal quantum correlations (almost) uniquely identify the state and measurements under consideration, a phenomenon known as \emph{self-testing}.
For instance the maximal violation of the Clauser-Horne-Shimony-Holt (CHSH) \cite{clauser69} inequality necessarily implies that the two parties share a singlet (up to local unitaries). While this was already pointed out by Popescu and Rohrlich in 1992 \cite{popescu92}, it was not widely known until the works of Mayers and Yao~\cite{mayers98, mayers04}. Since then self-testing has received substantial attention and led to the concept of \emph{device independence}, important in quantum cryptography~\cite{barrett05a, acin06, colbeck06, acin07, colbeck10} and beyond~\cite{scarani12}.

The central question in self-testing is: given a conditional probability distribution arising from measuring a (multipartite) quantum state, what can be deduced about the state and/or the measurements?

Most previous research has focused on the problem of certifying the quantum state shared between the devices. Among others we can self-test the singlet~\cite{mckague12}, graph states~\cite{mckague14}, high-dimensional maximally entangled states~\cite{slofstra11, yang13} or non-maximally entangled states of two qubits \cite{bamps15}. The common feature of these results is that the robustness is extremely weak, i.e.~we can only make a non-trivial statement if the observed statistics are $\varepsilon$-close to the ideal case (for $\varepsilon \approx 10^{-4}$). Self-testing statements of practically relevant robustness turn out to be significantly harder to prove and are currently restricted to a single analytic result~\cite{bardyn09} and one numerical technique known as the ``swap trick''~\cite{yang14, bancal15}. The ``swap trick'' relies on explicitly constructing a circuit extracting the desired state into an extra register and using the hierarchy for quantum correlations~\cite{navascues07, doherty08} to place a lower bound on the resulting fidelity. This is a versatile tool for obtaining robust self-testing statements for various entangled states but since the computational cost grows rapidly with dimensionality, so far it has only been applied to quantum systems of small dimensions (two qubits/qutrits~\cite{yang14, bancal15} or three/four qubits~\cite{wu14, pal14}).

Self-testing of measurements has received significantly less attention. Popescu and Rohrlich 
showed that the CHSH inequality is violated maximally only if the observables anticommute (which corresponds to maximal incompatibility)~\cite{popescu92}. McKague and Mosca showed how to certify more than two binary observables~\cite{mckague11},
Miller and Shi investigated which Bell inequalities are well-suited for self-testing~\cite{miller12}, Bamps and Pironio showed that non-maximally anticommuting observables can be self-tested using the tilted CHSH inequality~\cite{bamps15}, while \v{S}upi{\'c} et al.~recently showed that all measurements lying in a single plane of the Bloch sphere can be self-tested through chained inequalities~\cite{supic16}. To the best of our knowledge nothing is known about certifying measurements with more than two outcomes.

Let us also mention some recent works on self-testing which present a slightly different focus, e.g.~self-testing in parallel~\cite{mckague16, wu16}, producing a complete list of self-tests (within a particular Bell scenario)~\cite{wang16} or analysing semi-device independent scenarios~\cite{goh16, gheorghiu15, supic16a}.

In this Letter we present a new technique for proving analytic self-testing statements for quantum states. In the simplest case of self-testing the singlet using the CHSH inequality, we obtain a linear bound which improves on all the previously known results. We also consider self-testing the tripartite Greenberger-Horne-Zeilinger (GHZ) state~\cite{greenberger89} using the Mermin inequality~\cite{mermin90} which yields the first tight self-testing statement (``tight'' in the sense explained below, which is unrelated to the facet character of the corresponding Bell inequality). Our technique hinges on the idea that measurement operators can be used to construct local extraction maps. This gives rise to a family of operators and placing a lower bound on the spectrum of these operators immediately yields a self-testing statement. The technique can be straightforwardly applied to any Werner-Wolf-{\.Z}ukowski-Brukner inequality~\cite{werner01, zukowski02} (all Bell scenarios with two-settings and two outcomes per party) and we believe it is also applicable to more general scenarios.

\prlparagraph{Methods}Suppose that two parties, usually referred to as {\alice} and {\bob}, share some quantum state. If they had access to trusted measurement devices, they could perform tomography to deduce precisely what state they share. However we consider a more restrictive scenario in which {\alice} and {\bob} only have access to untrusted measurement devices. In other words their actions are limited to choosing the measurement setting and observing the outcome and hence the only information available to them is the conditional probability distribution (i.e.~the probability of observing outputs $a, b$ for inputs $x, y$). In this case one cannot hope to exactly identify the state shared by the devices: the two inherent limitations of self-testing are the inability to see local unitaries and the inability to detect auxiliary systems (on which the measurements act trivially). To formalise the problem we must therefore generalise the notion of {\alice} and {\bob} \emph{sharing} a particular state. A natural solution is to require that they are capable of locally (without communication) \emph{extracting} the desired state, which leads directly to a quantitative measure first proposed by Bardyn et al.~\cite{bardyn09}.
Let the \emph{target state} $\ket{\Psi}_{A'B'}$ be an arbitrary bipartite pure state (we assume all the subsystems to be finite-dimensional), $\Psi_{A'B'} = \ketbraq{\Psi}_{A'B'}$ be the corresponding density matrix and $F(\rho, \sigma) = \norm{ \sqrt{\rho} \sqrt{\sigma} }_{1}^{2}$ be the fidelity ($\norm{\cdot}_{1}$ is the trace norm). For an arbitrary bipartite \emph{input state} $\rho_{AB}$ we define the \emph{extractability} of $\Psi_{A'B'}$ from $\rho_{AB}$ as
\begin{equation*}
\Xi(\rho_{AB} \to \Psi_{A'B'}) := \max_{ \Lambda_{A}, \Lambda_{B} } F \big( ( \Lambda_{A} \otimes \Lambda_{B} ) (\rho_{AB}), \Psi_{A'B'} \big),
\end{equation*}
where the maximum is taken over all quantum channels (completely positive trace-preserving maps) of appropriate input and output registers.
Note that this is equivalent to first adding local ancillary registers (in an arbitrary state) and then performing local unitaries which extract the desired state into these registers. The extractability is convex in the input state and invariant under local unitaries (applied to either the input or the target state). Note that {\alice} and {\bob} can choose to discard their shares and replace them with some fixed states, which transforms $\rho_{AB}$ into an arbitrary product state. In particular this could be the product state corresponding to the largest Schmidt coefficient of $\ket{\Psi_{A'B'}}$ denoted by $\lmax$, which implies that $\Xi(\rho_{AB} \to \Psi_{A'B'}) \geq \lmax^{2}$ (in fact this turns out to be optimal whenever $\rho_{AB}$ is separable~\cite{shimony95}). Computing the extractability for an arbitrary pair of states seems to be a hard optimisation problem because the set of product channels is not convex.
This is not a major obstacle, since we do not intend to study the quantity itself but to investigate the trade-off between extractability and nonlocality. It is worth pointing out that finding the maximal violation of a fixed Bell inequality for a particular state is (for similar reasons) also believed to be hard~\cite{liang07}, which suggests that the two problems might be closely related.

In a self-testing problem we are given a target state $\Psi_{A'B'}$ and a Bell inequality $\cB$. Let $\beta_{C}$ and $\beta_{Q}$ be the maximal values of the inequality $\cB$ achieved within the classical and quantum theories respectively and for simplicity we assume that $\Psi_{A'B'}$ achieves the maximal quantum violation. The extractability-violation trade-off is captured by a function $\trsym_{\Psi, \cB} : [\beta_{C}, \beta_{Q}] \to [0, 1]$ defined as
\begin{equation*}
\trsym_{\Psi, \cB}(\beta) := \inf_{\rho_{AB} \in \cS_{\cB}(\beta)} \Xi(\rho_{AB} \to \Psi_{A'B'}),
\end{equation*}
where $\cS_{\cB}(\beta)$ is the set of bipartite states (of arbitrary dimension) that achieve the value of (at least) $\beta$ on the inequality $\cB$. This formulation is convenient as it (trivially) implies that an observed violation of $\beta$ guarantees that the shared state $\rho_{AB}$ satisfies
\begin{equation*}
\Xi(\rho_{AB} \to \Psi_{A'B'}) \geq \trsym_{\Psi, \cB}(\beta),
\end{equation*}
which is precisely a self-testing statement.

The lower bound on the extractability (based on the Schmidt decomposition of $\Psi_{A'B'}$) implies a trivial lower bound on the trade-off function: $\trsym_{\Psi, \cB}(\beta) \geq \lmax$ (trivial in the sense that it exhibits no $\beta$-dependence).
To derive an upper bound on $\trsym_{\Psi, \cB}(\beta)$ for a particular value of $\beta$ we write
\begin{equation}
\label{eq:mixture-violation}
\beta = p \beta_{Q} + (1 - p) \beta_{C}
\end{equation}
for some $p \in [0, 1]$. Then we consider the state
\begin{equation}
\label{eq:mixture}
\rho_{XYAB} := p \ketbraq{00}_{XY} \otimes \Psi_{AB} + (1-p) \ketbraq{11}_{XY} \otimes \sigma_{AB}
\end{equation}
where $\Psi_{AB}$ is the target state (we choose the dimensions of the registers $A, B$ to be the same as $A', B'$), $\sigma_{AB}$ is an arbitrary separable state and we consider the $XA | YB$ partition. By construction this state achieves the violation of $\beta$ so it suffices to place an upper bound on the extractability. We first use convexity and then apply the trivial bound of unity for $\ketbraq{00}_{XY} \otimes \Psi_{AB}$ and the separable bound $\lmax$ for $\ketbraq{11}_{XY} \otimes \sigma_{AB}$ to obtain
\begin{equation}
\label{eq:mixture-extractability}
\Xi(\rho_{XYAB} \to \Psi_{AB}) \leq p + (1 - p) \lmax.
\end{equation}
Combining Eqs.~\eqref{eq:mixture-violation} and \eqref{eq:mixture-extractability} leads to
\begin{equation}
\label{eq:upper-bound}
\trsym_{\Psi, \cB}(\beta) \leq \lmax + (1 - \lmax) \cdot \frac{\beta - \beta_{C}}{\beta_{Q} - \beta_{C}}.
\end{equation}

While the definition of extractability generalises to any number of parties, upper and lower bounds do not follow trivially. For instance the optimal $\beta$-independent lower bound for a multipartite state (sometimes referred to as the \emph{entanglement eigenvalue}) is only known for some special classes of states ~\cite{barnum01, wei03} (the general problem is equivalent  to computing the spectral norm of the corresponding tensor, which is known to be NP-hard~\cite{hillar13}). Often non-trivial bounds can be obtained by grouping parties and thus reducing it to a bipartite scenario.

\prlparagraph{Self-testing from operator inequalities}The only analytic self-testing result with practically relevant robustness is due to Bardyn et al.~\cite{bardyn09} and exploits the fact that local observables can be used to determine local unitaries that should be applied to each subsystem in order to extract the singlet. In the case of two binary observables per party we can use Jordan's lemma (for a precise statement and a simple proof we refer the reader to Oded Regev's lecture notes~\cite{regev06}) to write the observables in a block-diagonal form with blocks of size at most $2 \times 2$. After the unitary correction each (non-trivial) block is fully characterised by the angle between the observables. The goal is to show that a high CHSH violation implies high fidelity of the rotated state with the singlet and the main challenge is to derive a bound which is \emph{uniform}, i.e.~it holds \emph{for all angles} between the observables of {\alice} and {\bob}. (If the angles were publicly announced, then {\alice} and {\bob} would effectively share a two-qubit state for which better bounds have been proven~\cite{bardyn09}.)
In the current work we use these ideas to develop a new method for proving analytic self-testing bounds.

Our goal is to self-test the target state $\Psi_{A'B'}$ using a particular Bell inequality $\cB$ and here we show how to obtain linear self-testing statements of the form
\begin{equation}
\label{eq:linear-bound}
\trsym_{\Psi, \cB}(\beta) \geq s \beta + \mu
\end{equation}
for some real parameters $s, \mu \in \amsbb{R}$. The Bell operator of $\cB$ is defined as
\begin{equation*}
W: = \sum_{xyab} c^{xy}_{ab} P^{x}_{a} \otimes Q^{y}_{b},
\end{equation*}
where $c^{xy}_{ab} \in \amsbb{R}$ are real coefficients and
$\{P^{x}_{a} \}_{a}$ is the measurement performed by {\alice} on input $x$ (and similarly for {\bob}). Let $\Lambda_{A}$ and $\Lambda_{B}$ be local extraction channels constructed from the local observables.
Since the fidelity with a pure state can be written as the Hilbert-Schmidt inner product, we have
\begin{equation*}
F \big( ( \Lambda_{A} \otimes \Lambda_{B} ) (\rho_{AB}), \Psi_{A'B'} \big) = \ave{ (\Lambda_{A} \otimes \Lambda_{B} ) (\rho_{AB}), \Psi_{A'B'} }.
\end{equation*}
For every map $\Lambda$, there exists the dual map $\Lambda^{\dagger}$, which satisfies $\ave{ \Lambda(X), Y } = \ave{ X, \Lambda^{\dagger}(Y) }$ for all linear operators $X, Y$.
Thus we can rewrite the fidelity as $\tr (K \rho_{AB})$ for
\begin{equation*}
K := ( \Lambda_{A}^{\dagger} \otimes \Lambda_{B}^{\dagger} ) (\Psi_{A'B'}).
\end{equation*}
For a real constant $s > 0$ (to be chosen later) consider the operator $K - sW$ and suppose that $\mu \in \amsbb{R}$ is a lower bound on its spectrum or equivalently that the operator inequality
\begin{equation}
\label{eq:operator-inequality}
K \geq s W + \mu \mathbb{1}
\end{equation}
holds. Then computing the trace of this inequality with $\rho_{AB}$ leads directly to
\begin{equation}
\label{eq:fidelity-bound}
F \big( ( \Lambda_{A} \otimes \Lambda_{B} ) (\rho_{AB}), \Psi_{A'B'} \big) \geq s \beta + \mu.
\end{equation}
Proving the operator inequality~\eqref{eq:operator-inequality} for a \emph{particular choice} of measurement operators implies that the bound~\eqref{eq:fidelity-bound} holds for all states $\rho_{AB}$ for \emph{that particular choice} of measurement operators. Proving that the operator inequality~\eqref{eq:operator-inequality} holds \emph{for all} possible measurement operators (of arbitrary dimension) implies that the inequality~\eqref{eq:fidelity-bound} holds \emph{for all} quantum setups, which is precisely the meaning of inequality~\eqref{eq:linear-bound}. Since the measurement operators might be of arbitrary dimension, the operator $K - sW$ is difficult to analyse in general. Fortunately the analysis simplifies significantly whenever Jordan's lemma can be applied.

\prlparagraph{Qubit extraction maps}Without loss of generality we can assume that the measurement operators of {\alice} and {\bob} are projective and that all the Jordan blocks are non-trivial (see Appendix~\ref{app:sufficiency} for details).
We propose extraction maps that respect the block structure of the observables (i.e.~we only consider qubit-to-qubit channels) and we use identical maps for each party. In the first step we rotate each two-dimensional block so that the observables of {\alice}
can be written as
\begin{equation}
\label{eq:observables}
A_{r} = \cos a \cdot \sigma_{x} + (-1)^{r} \sin a \cdot \sigma_{z}
\end{equation}
for $r \in \{0, 1\}$ and some $a \in [0, \pi/2]$.
The observables of {\bob} are defined in the same manner with the angle denoted by $b \in [0, \pi/2]$. It is crucial to realise that this covers \emph{all possible choices} of observables. Thus if the operator inequality~\eqref{eq:operator-inequality} holds for qubit observables for all pairs $(a, b)$, then it also holds for arbitrary observables (see Appendix~\ref{app:sufficiency} for details). The second part of the extraction map is a dephasing channel
\begin{equation}
\label{eq:extraction-map}
[\Lambda(x)] (\rho) = \frac{ 1 + g(x) }{2} \, \rho + \frac{ 1 - g(x) }{2} \, \Gamma(x) \, \rho \, \Gamma(x),
\end{equation}
where $x$ is the angle (i.e.~$x = a$ for {\alice} and $x = b$ for {\bob}), $g(x) = ( 1 + \q ) ( \sin x + \cos x - 1 )$ and
\begin{align*}
\Gamma(x) &=
\begin{cases}
\sigma_{x} &\nbox{for} x \in [0, \pi/4],\\
\sigma_{z} &\nbox{for} x \in (\pi/4, \pi/2].
\end{cases}
\end{align*}
It is easy to check that $g(0) = g(\pi/2) = 0$ (full dephasing for compatible observables) and $g(\pi/4) = 1$ (no dephasing for maximally incompatible observables).
Having explicitly defined the extraction maps, let us now assess their performance in two concrete self-testing scenarios.

\prlparagraph{The CHSH inequality}The CHSH operator reads
\begin{equation*}
W(a, b) = \sum_{j,k \in \{0, 1\}} (-1)^{jk} A_{j} \otimes B_{k}.
\end{equation*}
The optimal violation of $\beta_{Q} = 2 \q$ is achieved only for $a = b = \pi/4$ and let us denote the optimal state (the eigenvector corresponding to the largest eigenvalue) by $\Phi_{AB}$ (equivalent to the singlet $(\ket{01} - \ket{10})/\q$ up to local unitaries). If {\alice} and {\bob} apply the extraction map~\eqref{eq:extraction-map}, we obtain the operator $K(a, b) = ( \Lambda(a) \otimes \Lambda(b) )( \Phi_{AB} )$ (the dephasing map is self-dual) for which we prove the following proposition.
\begin{prop-glob}
\label{prop:chsh}
Let $s = (4 + 5 \q)/16$ and $\mu = -(1 + 2 \q)/4$. Then the operator inequality
\begin{equation*}
K(a, b) \geq s W(a, b) + \mu \mathbb{1}
\end{equation*}
holds for all $a, b \in [0, \pi/2]$.
\end{prop-glob}
The proof of this proposition is one of the main technical contributions of this paper and can be found in Appendix~\ref{app:chsh}. As an immediate corollary we find that
\begin{equation}
\label{eq:chsh}
\trsym_{\Phi_{AB}, \cB_{\textnormal{CHSH}}}(\beta) \geq \frac{1}{2} + \frac{1}{2} \cdot \frac{\beta - \beta^{*}}{2 \q - \beta^{*}},
\end{equation}
where $\beta^{*} = \frac{ 16 + 14 \q }{ 17 } \approx 2.11$ is the threshold violation (for which the bound becomes non-trivial).
This result improves upon all previously known results and also follows closely the upper bound of Eq.~\eqref{eq:upper-bound} as shown in Fig.~\ref{fig:chsh}. As argued in Appendix~\ref{app:counterexample} the upper bound is unachievable in the interior $\beta \in (2, 2 \q)$, which might be related to the fact that the quantum value of the CHSH inequality does not reach its algebraic limit of $4$.
\begin{figure}
	\centering
	\includegraphics[scale=1]{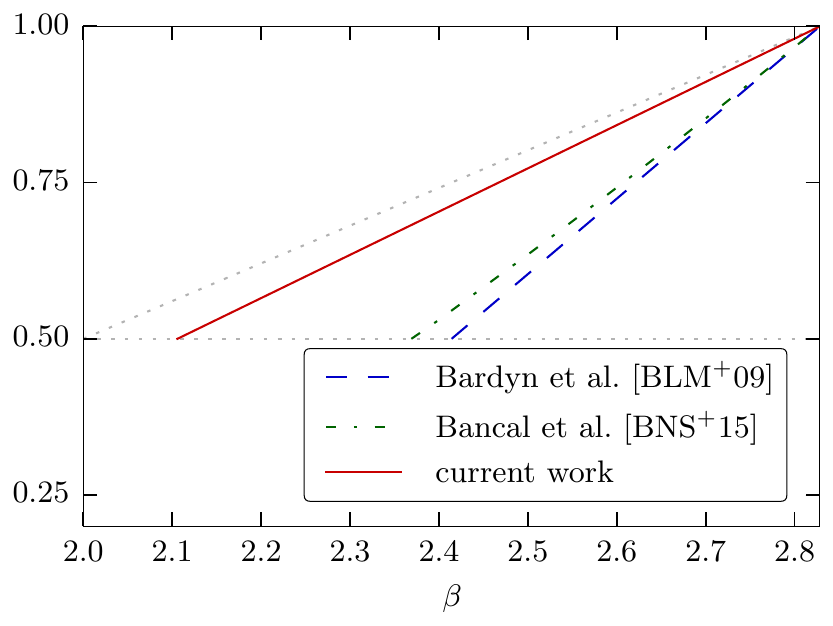}
	\caption{Comparison of lower bounds on $\trsym_{\Phi_{AB}, \cB_{\textnormal{CHSH}}}$. The gray dotted lines correspond to the trivial lower bound and the (unachievable) upper bound.}
	\label{fig:chsh}
\end{figure}
It was known previously that non-trivial fidelity with the singlet can be achieved if the violation exceeds $\approx 2.37$~\cite{bancal15}, which we currently improve to $\beta^{*} \approx 2.11$.

\begin{figure}[h!]
	\centering
	\includegraphics[scale=1]{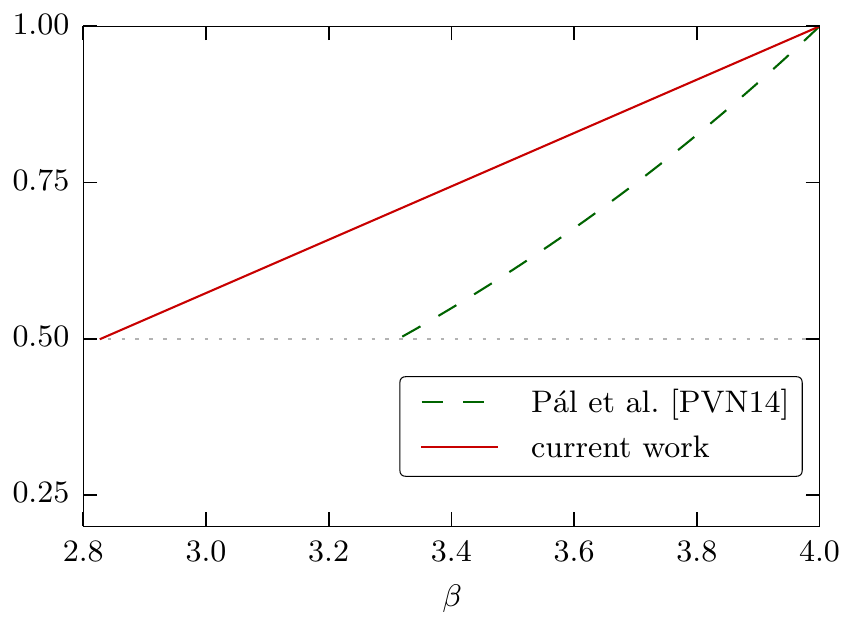}
	\caption{Comparison of the previously known lower bound with the exact value of $\trsym_{\Upsilon_{ABC}, \cB_{\textnormal{Mermin}}}$ derived in this work. The gray dotted line corresponds to the trivial lower bound.}
	\label{fig:mermin}
\end{figure}

\prlparagraph{The Mermin inequality}The Mermin operator reads
\begin{equation*}
W(a, b, c) = \sum_{j,k \in \{0, 1\}} (-1)^{jk} A_{j} \otimes B_{k} \otimes C_{j \oplus k}
\end{equation*}
and the observables of {\charlie} are defined analogous to Eq.~\eqref{eq:observables} with the angle denoted by $c \in [0, \pi/2]$.
The optimal quantum violation of $\beta_{Q} = 4$ is achieved only for $a = b = c = \pi/4$ and let us denote the corresponding state by $\Upsilon_{ABC}$ (equivalent to the tripartite GHZ state $(\ket{000} + \ket{111})/\q$ up to local unitaries). By considering any non-trivial bipartite cut it is easy to see that the optimal $\beta$-independent lower bound is $\trsym_{\Upsilon_{ABC}, \cB_{\textnormal{Mermin}}} \geq \frac{1}{2}$. To derive an upper bound we make two observations: (i) the violation up to $\gamma^{*} := 2 \q$ can be achieved by the state $\ket{\nu}_{ABC} = \ket{\Phi}_{AB} \ket{0}_{C}$ ({\alice} and {\bob} employ the CHSH strategy, while {\charlie} outputs a fixed outcome) and (ii) with respect to the $AB | C$ cut the state $\ket{\nu}_{ABC}$ is separable and the state $\Upsilon_{ABC}$ is equivalent to a singlet thus $\Xi(\nu_{ABC} \to \Upsilon_{A'B'C'}) = \frac{1}{2}$. From (i) and (ii) we immediately see that $\trsym_{\Upsilon_{ABC}, \cB_{\textnormal{Mermin}}}(\gamma) = \frac{1}{2}$ for $\gamma \in [2, \gamma^{*}]$. By considering mixtures of $\Upsilon_{ABC}$ and $\nu_{ABC}$ analogous to the state~\eqref{eq:mixture} we conclude that
\begin{equation}
\label{eq:mermin-upper-bound}
\trsym_{\Upsilon_{ABC}, \cB_{\textnormal{Mermin}}}(\gamma) \leq \frac{1}{2} + \frac{1}{2} \cdot \frac{\gamma - \gamma^{*}}{4 - \gamma^{*}}
\end{equation}
for $\gamma \in [\gamma^{*}, 4]$.
To derive a non-trivial lower bound we choose the same extraction map~\eqref{eq:extraction-map} for all three parties. This leads to $K(a, b, c) = ( \Lambda(a) \otimes \Lambda(b) \otimes \Lambda(c) )( \Upsilon_{ABC} )$ for which we prove the following proposition (see Appendix~\ref{app:mermin}).
\begin{prop-glob}
\label{prop:mermin}
Let $s = (2 + \q)/8$ and $\mu = - 1/\q$. Then the operator inequality
\begin{equation*}
K(a, b, c) \geq s W(a, b, c) + \mu \mathbb{1}
\end{equation*}
holds for all $a, b, c \in [0, \pi/2]$.
\end{prop-glob}
The resulting lower bound matches exactly the upper bound~\eqref{eq:mermin-upper-bound}, which implies that
\begin{equation}
\label{eq:mermin}
\trsym_{\Upsilon_{ABC}, \cB_{\textnormal{Mermin}}}(\gamma) = \frac{1}{2} + \frac{1}{2} \cdot \frac{\gamma - \gamma^{*}}{4 - \gamma^{*}}.
\end{equation}
It is worth pointing out that this constitutes the first self-testing statement which is provably tight. Comparison with the previously known bound is shown in Fig.~\ref{fig:mermin}.

\prlparagraph{Conclusions}We have presented a new technique for proving self-testing statements which relies on: (i) understanding how to construct extraction channels from measurement operators and (ii) analysing the resulting operator. We construct qubit extraction maps from two binary observables and derive analytic self-testing bounds for the CHSH and Mermin inequalities, which improve on previously known results.

To demonstrate the importance of these improvements consider the recent loophole-free Bell experiments~\cite{hensen15, hensen16}, which report the CHSH value of $\beta \approx 2.4$. In this case the previous results yield the singlet fidelity of (at least) $0.53$~\cite{bancal15}, which only slightly exceeds the trivial value of $\frac{1}{2}$. In contrast our results guarantee that the singlet fidelity is at least $0.70$, which constitutes a significant improvement. This can be used to obtain a lower bound on the distillable entanglement of the unknown state: we first perform local extraction and then distil entanglement from the resulting two-qubit state using standard procedures~\cite{dur07}.

An immediate follow-up problem is to certify the $n$-partite GHZ state using one of the Werner-Wolf-{\.Z}ukowski-Brukner inequalities~\cite{werner01, zukowski02}. For a fixed inequality our approach leads to an explicit family of operators and it suffices to place a lower bound on their spectrum, which makes the problem purely technical. Alternatively one could consider generalisations of the CHSH inequality~\cite{oppenheim10, slofstra11} with more than two settings per party. These inequalities might lead to robust self-testing of (bipartite) high-dimensional maximally entangled states. Finally one might attempt to construct extraction maps from measurements with more than two outcomes, e.g.~for the Collins-Gisin-Linden-Massar-Popescu inequalities~\cite{collins02}.

\prlparagraph{Acknowledgements}We would like to thank Jean-Daniel Bancal, Chris Majenz, Matthew McKague, Alexander M{\"u}ller-Hermes, Chris Perry and Marco Tomamichel for comments on an earlier version of this draft. We are grateful to Jean-Daniel Bancal and Tam{\'a}s V{\'e}rtesi for sharing their numerical data. We acknowledge stimulating discussions with Matthias Christandl, David Elkouss, Roberto Ferrara, Laura Man\v{c}inska, Marcin Paw{\l}owski, Filip Rozp\k{e}dek and Valerio Scarani. Most importantly we thank Alessandro Bruno for inspiring discussions on decoherence. We acknowledge financial support from the European Research Council (ERC Grant Agreement 337603), the Danish Council for Independent Research (Sapere Aude) and the Swiss National Science Foundation (Project PP00P2\_150734).
\bibliographystyle{alphaarxiv}
\bibliography{/home/jedrek/Projekty/tex/librarysan}
\appendix
\onecolumngrid
\section{Sufficiency of considering qubit observables}
\label{app:sufficiency}
We want to argue that if the operator inequality~\eqref{eq:operator-inequality} holds for qubit observables for all angles then it holds for all binary observables. For simplicity we sketch out the argument for the bipartite scenario but the multipartite generalisation is straightforward.

First note that we can without loss of generality assume that the measurements are projective (if they were not {\alice} and {\bob} would append local ancillas to make them projective). To avoid dealing with a direct sum of blocks of different size, for every trivial $1 \times 1$ block in the Jordan decomposition we add an extra dimension to the local Hilbert space. This simply corresponds to embedding the state in a larger Hilbert space (the state is not supported on these extra dimensions) but we can use these extra dimensions to turn every trivial block into some non-trivial block (the choice of the angle is irrelevant but for definiteness we can choose $a = b = 0$). Applying the same procedure on both sides ensures that we can write the operator $K$ as
\begin{equation*}
K = \sum_{xy} \ketbraq{x}_{X} \otimes \ketbraq{y}_{Y} \otimes K(a_{x}, b_{y}),
\end{equation*}
where $X$ and $Y$ are classical registers storing the block information of {\alice} and {\bob} respectively, $a_{x}, b_{y}$ are the angles and $K(a_{x}, b_{y})$ is a $4 \times 4$ operator corresponding to qubit observables. Similarly the Bell operator $W$ can be written as
\begin{equation*}
W = \sum_{xy} \ketbraq{x}_{X} \otimes \ketbraq{y}_{Y} \otimes W(a_{x}, b_{y}).
\end{equation*}
If the operator inequality
\begin{equation*}
K(a, b) \geq s W(a, b) + \mu \mathbb{1}
\end{equation*}
holds for qubit observables for all angles $a, b \in [0, \pi/2]$, it immediately implies that
\begin{equation*}
K \geq s W + \mu \mathbb{1}
\end{equation*}
holds for all observables.
\section{Operator inequalities for qubit observables}
Propositions~\ref{prop:chsh} and~\ref{prop:mermin} constitute the main technical contributions of this work. The proofs rely exclusively on elementary linear algebra and analysis but the actual calculations are rather lengthy.

Let us start by outlining the general proof technique. We aim to show that for some specific values of $s$ and $\mu$ the operator $T := K - s W - \mu \mathbb{1}$ is positive semidefinite for all angles between the local observables. In the first step we show that $T$ admits a generic block-diagonalisation into two-dimensional blocks (generic in the sense that these subspaces do not depend on the angles). More concretely we propose a partition of the identity into two-dimensional projectors $\sum_{x} P_{x} = \mathbb{1}$ which commute with $T$ ($[T, P_{x}] = 0$) for all $x$ and for all angles. Therefore
\begin{equation*}
T = \sum_{x} P_{x} T P_{x}
\end{equation*}
and it suffices to prove positivity of each block $M_{x} := P_{x} T P_{x}$. Since the rank of $M_{x}$ is at most $2$ we have
\begin{equation*}
M_{x} \geq 0 \quad \iff \quad \tr M_{x} \geq 0 \hspace{4pt} \wedge \hspace{3pt} (\tr M_{x})^{2} - \tr M_{x}^{2} \geq 0.
\end{equation*}
\subsection{Proof of Proposition~\ref{prop:chsh}}
\label{app:chsh}
The maximally entangled state $\Phi_{AB}$ is
\begin{equation*}
\Phi_{AB} = \frac{1}{4} \Big( \mathbb{1} \otimes \mathbb{1} + \sigma_{y} \otimes \sigma_{y} + \frac{1}{\q} \big[ \sigma_{x} \otimes \sigma_{x} + \sigma_{z} \otimes \sigma_{x} + \sigma_{x} \otimes \sigma_{z} - \sigma_{z} \otimes \sigma_{z} \big] \Big),
\end{equation*}
the dephased operator is $K(a, b) = \big( \Lambda_{A}(a) \otimes \Lambda_{B}(b) \big) ( \Phi_{AB} )$ and the Bell operator is
\begin{equation*}
W(a, b) = 2 \big( \cos a \cos b \cdot \sigma_{x} \otimes \sigma_{x} + \sin a \cos b \cdot \sigma_{z} \otimes \sigma_{x} + \cos a \sin b \cdot \sigma_{x} \otimes \sigma_{z} - \sin a \sin b \cdot \sigma_{z} \otimes \sigma_{z} \big).
\end{equation*}
Our goal is to show that the operator $T(a, b) = K(a, b) - s W(a, b) - \mu \mathbb{1}$ is positive semidefinite for
\begin{equation*}
s = \frac{4 + 5 \q}{16} \nbox{and} \mu = - \frac{1 + 2 \q}{4}
\end{equation*}
for all $a, b \in [0, \pi/2]$. It is easy to check that
\begin{equation*}
T(a, b) = ( H \otimes \sigma_{x} ) T(\pi/2 - a, b) ( H \otimes \sigma_{x} ),
\end{equation*}
which means that if we only care about the positivity of the operator it suffices to consider the range $a \in [0, \pi/4]$. By symmetry the same argument works for $b$ so from now on we assume that $a, b \in [0, \pi/4]$. Then the dephasing occurs in the Hadamard basis and we have
\begin{equation*}
K(a, b) = \frac{1}{4} \Big( \mathbb{1} \otimes \mathbb{1} + g(a) g(b) \sigma_{y} \otimes \sigma_{y} + \frac{1}{\q} \big[ \sigma_{x} \otimes \sigma_{x} + g(a) \sigma_{z} \otimes \sigma_{x} + g(b) \sigma_{x} \otimes \sigma_{z} - g(a) g(b) \sigma_{z} \otimes \sigma_{z} \big] \Big).
\end{equation*}
Writing $T(a, b)$ out gives
\begin{align*}
T(a, b) = &\bigg( \frac{1}{4} - \mu \bigg) \mathbb{1} \otimes \mathbb{1} + \frac{1}{4} g(a) g(b) \sigma_{y} \otimes \sigma_{y} + \frac{1}{4 \q} \bigg( \big[ 1 - 8 s \q \cos a \cos b \big] \sigma_{x} \otimes \sigma_{x} + \big[ g(a) - 8 s \q \sin a \cos b \big] \sigma_{z} \otimes \sigma_{x}\\
&+ \big[ g(b) - 8 s \q \cos a \sin b \big] \sigma_{x} \otimes \sigma_{z} - \big[ g(a) g(b) - 8 s \q \sin a \sin b\big] \sigma_{z} \otimes \sigma_{z} \bigg).
\end{align*}
Noticing that $[T(a, b), \sigma_{y} \otimes \sigma_{y}] = 0$ for all $a, b \in [0, \pi/2]$ leads us to consider projectors
\begin{equation*}
P_{x} := \frac{1}{2} \big( \mathbb{1} \otimes \mathbb{1} + (-1)^{x} \sigma_{y} \otimes \sigma_{y} \big)
\end{equation*}
for $x \in \{0, 1\}$. Then
\begin{equation*}
\tr M_{x} = \tr \big( P_{x} T(a, b) \big) = \frac{1}{2} - 2 \mu + (-1)^{x} \frac{ g(a) g(b) }{2}
\end{equation*}
and clearly $\tr M_{x} \geq 0$ (because $\mu \leq 0$ and $g(a), g(b) \in [0, 1]$). Computing $\tr M_{x}^{2}$ is slightly more involved but ultimately leads to
\begin{align*}
\tr M_{x}^{2} = \tr \big( P_{x} [T(a, b)]^{2} \big) &= 2 \bigg[ \bigg( \frac{1}{4} - \mu \bigg)^{2} + \frac{ 1 + g(a)^{2} + g(b)^{2} + 3 \big[ g(a) g(b) \big]^{2} }{32} + 4 s^{2} - \frac{s}{\q} ( \cos a + g(a) \sin a ) ( \cos b + g(b) \sin b) \bigg]\\
&+ \frac{(-1)^{x}}{2} \bigg[ ( 1 - 2 \mu ) g(a) g(b) - 2 s \q \big[ g(a) \cos a + \sin a \big] \big[ g(b) \cos b + \sin b \big] + 16 s^{2} \sin 2a \sin 2b \bigg].
\end{align*}
Our goal is to show the positivity of $\lambda_{x}(a, b) := (\tr M_{x})^{2} - \tr M_{x}^{2}$ for $x \in \{0, 1\}$ and $a, b \in [0, \pi/4]$. Writing the entire expression out gives
\begin{align*}
\lambda_{x}(a, b) &= 2 \bigg( \frac{1}{4} - \mu \bigg)^{2} + \frac{ -1 - g(a)^{2} - g(b)^{2} + \big[ g(a) g(b) \big]^{2} }{16} - 8 s^{2} + s \q ( \cos a + g(a) \sin a ) ( \cos b + g(b) \sin b)\\
&+ (-1)^{x} \bigg[ - \mu g(a) g(b) + s \q \big[ g(a) \cos a + \sin a \big] \big[ g(b) \cos b + \sin b \big] - 8 s^{2} \sin 2a \sin 2b \bigg].
\end{align*}
It is convenient to introduce new coordinates: $u = ( a + b )/2$ and $t = \cos [ ( a - b )/2 ]$. The domain $(a, b) \in [0, \pi/4]$ maps onto $u \in [0, \pi/4]$ and
\begin{equation*}
t \in  \big[ \cos \big( \min \{ u, \pi/4 - u \} \big), 1 \big]
\end{equation*}
but for simplicity we extend it to $ t \in [\zeta, 1]$ for $\zeta = \cos(\pi/8)$.
\begin{itemize}
\item For $x = 0$ we have
\begin{align*}
\lambda_{0}(u, t) &= \frac{15 + 12\q}{2} t^{4} - 2 (10 + 7\q) (\sin u + \cos u) t^{3}  + \bigg( \frac{27 + 17\q}{2} + \frac{27 + 19\q}{2} \sin 2u \bigg) t^{2}\\
&- \bigg( (3 + 2\q) \sin u - \frac{2 + \q}{2} \cos u - \frac{8 + 5\q}{2} ( \sin u - \cos u) (\sin u)^{2} \bigg) t\\
&- \frac{4 + 3 \q}{2} \sin 2u + \frac{2 + \q}{4} \cos 4u - \frac{10 + 5\q}{4}.
\end{align*}
To prove $\lambda_{0}(u, t) \geq 0$ for $u \in [0, \pi/4]$ and $t \in [ \zeta, 1]$ we lowerbound it by a quadratic function. First we check that $\lambda_{0}(u, \zeta) \geq \lambda_{0}(u, 1) \geq 0$ (for all $u \in [0, \pi/4]$), which implies that the quadratic function
\begin{equation*}
q(u, t) := \big[ \lambda_{0}(u, \zeta) - \lambda_{0},(u, 1) \big] \bigg( \frac{1 - t}{1 - \zeta} \bigg)^{2} + \lambda_{0}(u, 1)
\end{equation*}
is non-negative. The final step is to verify that the difference $h(u, t) := \lambda_{0}(u, t) - q(u, t)$ is non-negative $h(u, t) \geq 0$, which follows directly from vanishing on the boundary ($h(u, \zeta) = h(u, 1) = 0$) and concavity (check that $\partial^{2} h / \partial t^{2} \leq 0$).
\item For $x = 1$ we have
\begin{align*}
\lambda_{1}(u, t) &= \frac{25 + 16\q}{2} t^{4} - \frac{1}{2} \big[ (8 + 5\q) \sin u + (26 + 19\q) \cos u \big] t^{3}\\
&+ \frac{1}{4} \big[ - (38 + 24\q) + (31 + 22\q) \sin 2u - (23 + 16\q) \cos 2u  \big] t^{2}\\
&+ \frac{1}{2} \big[ - (2 + 2\q) \sin u + (33 + 24\q) \cos u - (40 + 28\q) (\sin u)^{2} \cos u - (6 +4\q) (\sin u)^{3} \big] t\\
&- \frac{19 + 14\q}{8} \sin 2u + \frac{5 + 2\q}{8} \cos 2u - \frac{23 + 16\q}{16} \sin 4u - \frac{25 + 16\q}{16} \cos 4u + \frac{3 + 4\q}{16}.
\end{align*}
It is easy to verify that $\lambda_{1}(u, t)$ is convex in $t$ (check that $\partial^{2} \lambda_{1} / \partial t^{2} > 0$) which means it can be lowerbounded by linear functions tangent to it. Checking these at $t = \zeta$ and $t = 1$ suffices to prove positivity.
\end{itemize}
\subsection{Proof of Proposition~\ref{prop:mermin}}
\label{app:mermin}
The optimal state $\Upsilon_{ABC}$ is
\begin{align*}
\Upsilon_{ABC} &= \frac{1}{8} \big( \mathbb{1} \otimes \mathbb{1} \otimes \mathbb{1} + \sigma_{y} \otimes \sigma_{y} \otimes \mathbb{1} - \sigma_{y} \otimes \mathbb{1} \otimes \sigma_{y} - \mathbb{1} \otimes \sigma_{y} \otimes \sigma_{y} \big)\\
&+ \frac{1}{8 \q} \big( \sigma_{x} \otimes \sigma_{x} \otimes \sigma_{x} - \sigma_{x} \otimes \sigma_{x} \otimes \sigma_{z} + \sigma_{x} \otimes \sigma_{z} \otimes \sigma_{x} + \sigma_{z} \otimes \sigma_{x} \otimes \sigma_{x}\\
&+ \sigma_{x} \otimes \sigma_{z} \otimes \sigma_{z} + \sigma_{z} \otimes \sigma_{x} \otimes \sigma_{z} - \sigma_{z} \otimes \sigma_{z} \otimes \sigma_{x} + \sigma_{z} \otimes \sigma_{z} \otimes \sigma_{z} \big),
\end{align*}
the dephased operator is $K(a, b, c) = \big( \Lambda_{A}(a) \otimes \Lambda_{B}(b) \otimes \Lambda_{C}(c) \big) ( \Upsilon_{ABC} )$ and the Bell operator is
\begin{align*}
W &= 2 \big( \cos a \cos b \cos c \, \sigma_{x} \otimes \sigma_{x} \otimes \sigma_{x} - \cos a \cos b \sin c \, \sigma_{x} \otimes \sigma_{x} \otimes \sigma_{z} + \cos a \sin b \cos c \, \sigma_{x} \otimes \sigma_{z} \otimes \sigma_{x}\\
&+ \sin a \cos b \cos c \, \sigma_{z} \otimes \sigma_{x} \otimes \sigma_{x} + \cos a \sin b \sin c \, \sigma_{x} \otimes \sigma_{z} \otimes \sigma_{z} + \sin a \cos b \sin c \, \sigma_{z} \otimes \sigma_{x} \otimes \sigma_{z}\\
&- \sin a \sin b \cos c \, \sigma_{z} \otimes \sigma_{z} \otimes \sigma_{x} + \sin a \sin b \sin c \, \sigma_{z} \otimes \sigma_{z} \otimes \sigma_{z} \big).
\end{align*}
Our goal is to show that the operator $T(a, b, c) = K(a, b, c) - s W(a, b, c) - \mu \mathbb{1}$ is positive semidefinite for
\begin{equation*}
s = \frac{2 + \q}{8} \nbox{and} \mu = - \frac{1}{\q}
\end{equation*}
for all $a, b, c \in [0, \pi/2]$. It is easy to check that
\begin{align*}
T(a, b, c) &= (H \otimes \sigma_{x} \otimes \sigma_{x}) T(\pi/2 - a, b, c) (H \otimes \sigma_{x} \otimes \sigma_{x})\\
&= (\sigma_{x} \otimes H \otimes \sigma_{x}) T(a, \pi/2 - b, c) (\sigma_{x} \otimes H \otimes \sigma_{x})\\
&= (\sigma_{x} \otimes \sigma_{x} \otimes V) T(a, b, \pi/2 - c) (\sigma_{x} \otimes \sigma_{x} \otimes V),
\end{align*}
where $V = (\sigma_{x} - \sigma_{z})/\q$. Therefore it suffices to consider $a, b, c \in [0, \pi/4]$. Then the dephasing occurs in the Hadamard basis and we have
\begin{align*}
K(a, b, c) &= \frac{1}{8} \big( \mathbb{1} \otimes \mathbb{1} \otimes \mathbb{1} + g(a) g(b) \sigma_{y} \otimes \sigma_{y} \otimes \mathbb{1} - g(a) g(c) \sigma_{y} \otimes \mathbb{1} \otimes \sigma_{y} - g(b) g(c) \mathbb{1} \otimes \sigma_{y} \otimes \sigma_{y} \big)\\
&+ \frac{1}{8 \q} \big( \sigma_{x} \otimes \sigma_{x} \otimes \sigma_{x} - g(c) \sigma_{x} \otimes \sigma_{x} \otimes \sigma_{z} + g(b) \sigma_{x} \otimes \sigma_{z} \otimes \sigma_{x} + g(a) \sigma_{z} \otimes \sigma_{x} \otimes \sigma_{x}\\
&+ g(b) g(c) \sigma_{x} \otimes \sigma_{z} \otimes \sigma_{z} + g(a) g(c) \sigma_{z} \otimes \sigma_{x} \otimes \sigma_{z} - g(a) g(b) \sigma_{z} \otimes \sigma_{z} \otimes \sigma_{x} + g(a) g(b) g(c) \sigma_{z} \otimes \sigma_{z} \otimes \sigma_{z} \big).
\end{align*}
Noticing that $[T(a, b, c), \mathbb{1} \otimes \sigma_{y} \otimes \sigma_{y}] = [T(a, b, c), \sigma_{y} \otimes \mathbb{1} \otimes \sigma_{y} ] = [T(a, b, c), \sigma_{y} \otimes \sigma_{y} \otimes \mathbb{1}] = 0$ for all $a, b, c \in [0, \pi/2]$ leads us to consider projectors
\begin{equation*}
P_{x_{1} x_{2}} = \frac{1}{4} \big( \mathbb{1} \otimes \mathbb{1} \otimes \mathbb{1} + (-1)^{x_{1}} \sigma_{y} \otimes \sigma_{y} \otimes \mathbb{1} + (-1)^{x_{2}} \sigma_{y} \otimes \mathbb{1} \otimes \sigma_{y} + (-1)^{x_{1} + x_{2}} \mathbb{1} \otimes \sigma_{y} \otimes \sigma_{y} \big)
\end{equation*}
for $x_{1}, x_{2} \in \{0, 1\}$. It is easy to check that
\begin{equation*}
\tr M_{x_{1}, x_{2}} = \tr \big( P_{x_{1}, x_{2}} T(a, b, c) \big) = 2 \bigg( \frac{1}{8} - \mu \bigg) + \frac{1}{4} \big[ (-1)^{x_{1}} g(a) g(b) - (-1)^{x_{2}} g(a) g(c) - (-1)^{x_{1} + x_{2}} g(b) g(c) \big],
\end{equation*}
which is easily seen to be positive
\begin{equation*}
\tr M_{x_{1}, x_{2}} \geq 2 \bigg( \frac{1}{8} - \mu \bigg) - \frac{3}{4} \geq 0.
\end{equation*}
Computing $\tr M_{x_{1}, x_{2}}^{2}$ is a rather lengthy calculation so let us go directly to the final expression. Our goal is to show the positivity of $\lambda_{x_{1}, x_{2}}(a, b, c) := (\tr M_{x_{1}, x_{2}})^{2} - \tr M_{x_{1}, x_{2}}^{2}$ for $x_{1}, x_{2} \in \{0, 1\}$ and $a, b, c \in [0, \pi/4]$. Writing the entire expression out gives
\begin{align*}
\lambda_{x_{1} x_{2}}(a, b, c) &= 2 \bigg( \frac{1}{8} - \mu \bigg)^{2} - \frac{1}{32} + \frac{ [1 - g(a)^{2}][1 - g(b)^{2}][1 - g(c)^{2}]}{64}\\
&- 8 s^{2} + \frac{s}{\q} [\cos a + g(a) \sin a] [\cos b + g(b) \sin b] [\cos c + g(c) \sin c]\\
&+ (-1)^{x_{1}} \bigg[ - \frac{ \mu g(a) g(b) }{2} - 8 s^{2} \sin 2a \sin 2b + \frac{s}{\q} [ g(a) \cos a + \sin a ] [ g(b) \cos b + \sin b ] [\cos c + g(c) \sin c] \bigg]\\
&+ (-1)^{x_{2}} \bigg[ \frac{ \mu g(a) g(c) }{2} + 8 s^{2} \sin 2a \sin 2c - \frac{s}{\q} [ g(a) \cos a + \sin a ] [ \cos b + g(b) \sin b ] [ g(c) \cos c + \sin c] \bigg]\\
&+ (-1)^{x_{1} + x_{2}} \bigg[ \frac{ \mu g(b) g(c) }{2} + 8 s^{2} \sin 2b \sin 2c - \frac{s}{\q} [ \cos a + g(a) \sin a ] [ g(b) \cos b + \sin b ] [ g(c) \cos c + \sin c] \bigg].
\end{align*}
First we observe that the cases $(x_{1}, x_{2}) \in \{ (0, 0), (1, 0), (1, 1) \}$ are equivalent in the sense that
\begin{equation*}
\lambda_{00}(a, b, c) = \lambda_{10}(c, b, a) = \lambda_{11}(a, c, b).
\end{equation*}
Therefore it suffices to prove positivity of one of them (for all $a, b, c \in [0, \pi/4]$). The case of $x_{1} = 0, x_{2} = 1$ is qualitatively different and turns out to be strictly more restrictive then the other ones. To show this we first prove that
\begin{equation*}
\Delta \lambda(a, b, c) := \lambda_{00}(a, b, c) - \lambda_{01}(a, b, c) \geq 0
\end{equation*}
for all $a, b, c \in [0, \pi/4]$. Computing the difference gives
\begin{align*}
\Delta \lambda(a, b, c) &= \mu [ g(a) + g(b) ] g(c) + 16 s^{2} ( \sin 2a + \sin 2b ) \sin 2c\\
&- s \q [g(c) \cos c + \sin c] \Big( [ g(a) + g(b) ] \cos (a - b) + [1 + g(a) g(b)] \sin (a + b) \Big).
\end{align*}
Since the case of $c = 0$ is trivial (all the terms vanish), we divide through by $\sin 2c$. It is easy to verify that
\begin{gather*}
\frac{g(c)}{\sin 2c} \leq \frac{1 + \q}{2},\\
\frac{g(c) \cos c + \sin c}{ \sin 2c } \leq \frac{2 + \q}{2}.
\end{gather*}
Applying these inequalities and using the substitution of variables mentioned before ($u = ( a + b )/2$ and $t = \cos [ ( a - b )/2 ]$) gives
\begin{align*}
\frac{\Delta \lambda(u, t)}{2s \sin 2c} \geq f(u, t) := &- (6 + 4\q) (\sin u + \cos u) t^{3}+ \big[ 6 + 4\q + (1 - \q) \sin 2u \big] t^{2} + (\sin u + \cos u) \big[ 1 + (7 + 5\q) \sin 2u \big] t\\
& - \frac{9 + 5\q}{2} \sin 2u - \frac{7 + 5\q}{2} (\sin 2u)^{2} - 1.
\end{align*}
The final step is to check that $f(u, t)$ is concave in $t$ (check that $\partial^{2} f / \partial t^{2} \leq 0$
) and then verify that the function is non-negative on the boundaries ($f(u, \zeta) \geq 0$ and $f(u, 1) \geq 0$).

The last step is to prove that $\lambda_{01}(a, b, c) \geq 0$. In this case we use the substitution
\begin{align*}
x &= 1 - \sin (a + \pi/4),\\
y &= 1 - \sin (b + \pi/4),\\
z &= 1 - \sin (c + \pi/4),
\end{align*}
which maps the domain $a, b, c \in [0, \pi/4]$ onto $x, y, z \in [0, \eta] $ for $\eta := 1 - 1/\sqrt{2}$. This allows us to write
\begin{equation*}
\lambda_{01}(a, b, c, x, y, z) = P(x, y, z) + Q(a, b, c)
\end{equation*}
where
\begin{align*}
P(x, y, z) &:= (3 + 2\q) (xy + xz + yz) - \frac{4 + 3 \q}{2} ( x^{2} y + x^{2} z + x y^{2} + x z^{2} + y^{2} z + y z^{2})\\
&+ \frac{4 + 3 \q}{2} ( x^{2} y^{2} + x^{2} z^{2} + y^{2} z^{2}) + \frac{69 + 48 \q}{4} x y z ( x + y + z )\\
&- \frac{50 + 35 \q}{8} x y z ( x y + x z + y z ) - \frac{128 + 87 \q}{4} x y z - \frac{31 + 22\q}{8} x^{2} y^{2} z^{2}
\end{align*}
and
\begin{equation*}
Q(a, b, c) := \frac{2 + \q}{8} [ 1 - g(a) ] [ 1 - g(b) ] [ 1 - g(c) ] \cos ( a + \pi/4 ) \cos (b + \pi/4) \cos (c + \pi/4).
\end{equation*}
The second term is non-negative by inspection. To bound the polynomial term we first observe that
\begin{equation*}
x^{2} y + x^{2} z + x y^{2} + x z^{2} + y^{2} z + y z^{2} = (xy + xz + yz)(x + y +z ) - 3 x y z \leq 3 \eta (xy + xz + yz) - 3 x y z,
\end{equation*}
which implies
\begin{align*}
P(x, y, z) &\geq \frac{3 + \q}{2} (xy + xz + yz) + \frac{4 + 3 \q}{2} ( x^{2} y^{2} + x^{2} z^{2} + y^{2} z^{2}) + \frac{69 + 48 \q}{4} x y z ( x + y + z )\\
&- \frac{50 + 35 \q}{8} x y z ( x y + x z + y z ) - \frac{104 + 69 \q}{4} x y z - \frac{31 + 22\q}{8} x^{2} y^{2} z^{2}.
\end{align*}
In the final step we place a lower bound on this expression in terms of the geometric mean $r := \sqrt[3]{xyz}$. By the inequality of arithmetic and geometric means we have
\begin{gather*}
x + y + z \geq 3 r,\\
xy + yz + xz \geq 3 r^{2},\\
x^{2} y^{2} + x^{2} z^{2} + y^{2} z^{2} \geq 3 r^{4}.
\end{gather*}
Combining it with the trivial $xy + yz + xz \leq 3 \eta^{2}$ implies $P(x, y, z) \geq P'(r)$ for
\begin{equation*}
P'(r) := \frac{9 + 3\q}{2} r^{2} - \frac{446 + 291\q}{16} r^{3} + \frac{81\q}{2} r^{4} - \frac{31 + 22\q}{8} r^{6},
\end{equation*}
which is easily verified to be positive for $r \in [0, \eta]$.
\section{Counterexample to the upper bound}
\label{app:counterexample}
To show that the trivial upper bound is not achievable, we present a family of bipartite states that achieve the CHSH violation of $\beta \in (2, 2 \q)$ and argue that no extraction channels produce a singlet of fidelity reaching the upper bound~\eqref{eq:upper-bound}. The following argument could probably be extended to give us an improved upper bound but this is beyond the scope of this work.

We consider a family of states in which each party holds two qubits. On each side the first qubit (labelled by $X$ and $Y$ respectively) is used as a classical register whereas the remaining qubits (labelled by $A$ and $B$ respectively) contain a quantum state. Consider the family of states
\begin{equation*}
\rho_{XYAB} = \sum_{xy} p_{xy} \ketbraq{x}_{X} \otimes \ketbraq{y}_{Y} \otimes \tau_{AB}^{xy},
\end{equation*}
where $p_{11} = \nu$ and $p_{00} = p_{01} = p_{10} = (1 - \nu)/3$ for $\nu \in (0, 1)$ and the conditional states are
\begin{align*}
\tau_{AB}^{00} &= \tau_{AB}^{01} = \tau_{AB}^{10} = \frac{1}{4} ( \mathbb{1} \otimes \mathbb{1} + \sigma_{x} \otimes \sigma_{x} ),\\
\tau_{AB}^{11} &= \frac{1}{4} \big( \mathbb{1} \otimes \mathbb{1} + \frac{1}{\q} [ \sigma_{x} \otimes \sigma_{x} + \sigma_{x} \otimes \sigma_{z} + \sigma_{z} \otimes \sigma_{x} - \sigma_{z} \otimes \sigma_{z} ] + \sigma_{y} \otimes \sigma_{y} \big).
\end{align*}
The states $\tau_{AB}^{00}, \tau_{AB}^{01}, \tau_{AB}^{10}$ are perfectly (classically) correlated in the Hadamard basis, while $\tau_{AB}^{11}$ is a (pure) maximally entangled state. For simplicity we will assume that $\tau_{AB}^{11}$ is precisely the state that {\alice} and {\bob} are trying to extract. The observables of {\alice} and {\bob} are
\begin{align*}
A_{0} &= B_{0} = \ketbraq{0} \otimes \sigma_{x} + \ketbraq{1} \otimes \sigma_{x},\\
A_{1} &= B_{1} = \ketbraq{0} \otimes \sigma_{x} + \ketbraq{1} \otimes \sigma_{z}.
\end{align*}
It is easy to verify that the CHSH value equals
\begin{equation*}
\beta = 2 + (2 \q - 2) \nu
\end{equation*}
and the value of the upper bound equals $(1 + \nu)/2$. To reach it {\alice} and {\bob} would have to achieve the fidelity of $1$ for $x = y = 1$ and of $\frac{1}{2}$ for the remaining states. We show that this is not possible.

Since the register $X$ is classical, we can without loss of generality assume that the extraction channel applied by {\alice} is equivalent to a pair of qubit channels of the form $\{ \Lambda_{A \to A'}^{0}, \Lambda_{A \to A'}^{1} \}$ corresponding to different values of $x$ (and clearly an analogous argument holds for {\bob}). Therefore we have to optimise over two qubit channels for each party. In order to achieve the fidelity of $1$ for $x = y = 1$ {\alice} and {\bob} must either do nothing or apply unitaries which leave the state $\tau_{AB}^{11}$ unchanged. Let us for now assume that they do nothing, i.e.~that $\Lambda_{A \to A'}^{1}$ and $\Lambda_{B \to B'}^{1}$ are the identity channel. Once we know that $\Lambda_{B \to B'}^{1}$ is the identity channel, the only manner to achieve the optimal fidelity for $x = 0, y = 1$ is for {\alice} to apply a unital channel which satisfies
\begin{equation}
\label{eq:channel-constraint}
\Lambda_{A \to A'}^{0} ( \sigma_{x} ) = \frac{ \sigma_{x} + \sigma_{z} }{\q}.
\end{equation}
By symmetry considering the case of $x = 1, y = 0$ leads to the same conclusion for {\bob}. This determines all extraction channels and we can check that for $x = y = 0$ {\alice} and {\bob} end up with the state
\begin{equation*}
\frac{1}{4} \big( \mathbb{1} \otimes \mathbb{1} + \frac{1}{2} [ \sigma_{x} \otimes \sigma_{x} + \sigma_{x} \otimes \sigma_{z} + \sigma_{z} \otimes \sigma_{x} + \sigma_{z} \otimes \sigma_{z} ] \big),
\end{equation*}
whose fidelity with $\tau_{AB}^{11}$ equals $\frac{1}{4} + \frac{1}{4\q}$ which is less than the desired $\frac{1}{2}$.

The case of {\alice} and {\bob} applying some non-trivial unitaries for $x = y = 1$ turns out to be similar: once $\Lambda_{A \to A'}^{1}$ and $\Lambda_{B \to B'}^{1}$ are fixed, we obtain constraints similar to Eq.~\eqref{eq:channel-constraint} for $\Lambda_{A \to A'}^{0}$ and $\Lambda_{B \to B'}^{0}$. These constraints uniquely determine the state $( \Lambda_{A \to A'}^{0} \otimes \Lambda_{B \to B'}^{0} ) (\tau_{AB}^{00})$ and one can check that it always ends up being slightly ``misaligned'' with $\tau_{AB}^{11}$.
\end{document}